\documentclass{mathincs}
\def\Oo{{\mathcal O}}

 \newtheorem{thm}{Theorem}[section]

  \newtheorem{con}[thm]{Conjecture}
 \newtheorem{prop}[thm]{Proposition}
 \theoremstyle{definition}
 \newtheorem{defn}[thm]{Definition}
 \theoremstyle{remark}
 \newtheorem{rem}[thm]{Remark}
 \newtheorem*{ex}{Example}
 \numberwithin{equation}{section}

\def\res{\mathop{\rm res}\nolimits}
\def\disc{\mathop{\rm disc}\nolimits}
\def\R{\mathbb{R}}
\def\Z{\mathbb{Z}}

\def\C{\mathbb{C}}

\usepackage{hyperref}
\usepackage{enumerate}
\usepackage{bm}
\usepackage{color}
\usepackage{amsfonts}
\usepackage{todonotes}

\begin{document}

%
%

\title[Iterated Resultants and Rational Functions in Real QE]
 {Iterated Resultants and Rational Functions in\\Real Quantifier Elimination}
 
\author[J.H. Davenport]{James H. Davenport}
\address{%
University of Bath\\
Bath, BA2 7AY \\
United Kingdom}
\email{J.H.Davenport@bath.ac.uk}

\author[M. England]{Matthew England}
\address{%
Coventry University, \\
Coventry, CV1 5FB \\
United Kingdom}
\email{Matthew.England@coventry.ac.uk}

\author[S. McCallum]{Scott McCallum}
\address{%
Macquarie University, \\
Sydney, NSW 2109 \\
Australia}
\email{Scott.Mccallum@mq.edu.au}

\author[A.K. Uncu]{Ali K. Uncu}
\address{%
University of Bath\\
Bath, BA2 7AY \\
United Kingdom}
\email{aku21@bath.ac.uk\vspace{0.8mm}}

\address{%
Austrian Academy of Sciences\\
Johann Radon Institute for Computational and Applied Mathematics\\
Linz, 4040 \\
Austria}
\email{akuncu@ricam.oeaw.ac.at}



\date{\today}


\begin{abstract}
This paper builds and extends on the authors' previous work related to the algorithmic tool, Cylindrical Algebraic Decomposition (CAD), and one of its core applications, Real Quantifier Elimination (QE).  These topics are at the heart of symbolic computation and were first implemented in computer algebra systems decades ago, but have recently received renewed interest as part of the ongoing development of SMT solvers for non-linear real arithmetic.  

First, we consider the use of iterated univariate resultants in traditional CAD, and how this leads to inefficiencies, especially in the case of an input with multiple equational constraints.   We reproduce the workshop paper [Davenport \& England, 2023], adding important clarifications to our suggestions first made there to make use of multivariate resultants in the projection phase of CAD.  We then consider an alternative approach to this problem first documented in [McCallum \& Brown, 2009] which redefines the actual object under construction, albeit only in the case of two equational constraints.  We correct an unhelpful typo and provide a proof missing from that paper.  

We finish by revising the topic of how to deal with SMT or Real QE problems expressed using rational functions (as opposed to the usual polynomial ones) noting that these are often found in industrial applications.  We revisit a proposal made in [Uncu, Davenport and England, 2023] for doing this in the case of satisfiability, explaining why such an approach does not trivially extend to more complicated quantification structure and giving a suitable alternative.
\end{abstract}

\maketitle

\section{Introduction}

In this paper we will extend some work on Cylindrical Algebraic Decomposition (CAD) and Quantifier Elimination for logical formulae with polynomial constraints over the reals (Real QE).  These extensions take the form of tying up loose ends from three papers that (subsets of) the present authors were involved in.  Two of the three papers were published within the proceedings of the SC-Square Workshop \cite{DavenportEngland2023a}, \cite{Uncuetal2023a} and the other was published earlier on a related topic to the first \cite{McCallumBrown2009}.  All three extensions should be of interest to the SC-Square community, in particular those working on SMT in non-linear real arithmetic utilising elements of CAD theory, or on computer algebra for Real QE.

\subsection{Cylindrical algebraic decomposition}

CAD is an algorithm (and associated mathematical object) that decomposes real space relative to an input (usually a set of polynomials or a logical formula whose atoms are polynomial constraints).  CAD algorithms usually work through a process of projection (whereby additional projection polynomials are produced iteratively eliminating variables in each stage according to a fixed variable ordering) and lifting (where decompositions of $\mathbb{R}^i$ are iteratively produced for increasing $i$ relative to those projection polynomials in the first $i$ variables of the ordering).  

During lifting, each cell in $\R^i$ is lifted to a collection (\emph{stack}) of cells in $\R^{i+1}$ by evaluating the polynomials in $\R^{i+1}$ at the sample point in $\R^i$ and then using univariate real root isolation.  We are able to conclude the decomposition over the sample as representative of the decomposition over the whole cell so long as the polynomials were \emph{delineable} over the cell, which informally means they have the same root structure over any point in the cell (the same number of real roots in the same order). The reader is referred to the collection \cite{CavinessJohnson1998} for full details.  

Over the past decade the theory of CAD has been used and further developed for use in the context of SMT for non-linear real arithmetic, which is the special case of QE where all variables are existentially quantified.  CAD implementations have been developed that work more efficiently when used in this context \cite{KremerAbraham2019a}.  Further, whole new algorithms have been developed which use the CAD theory in new ways, e.g. NLSAT \cite{JovanovicdeMoura2012a}, NuCAD \cite{Brown2015a} and Cylindrical Algebraic Coverings \cite{Abrahametal2021a}, \cite{KremerNalbach2023a}, in part, due to the SC-Square initiative \cite{Abrahametal2016b}.  We expect that most of what we develop in this paper for CAD could be adapted for these other variants.

\subsection{Plan and contributions}

We define in Section \ref{sec:resultants} the notation and theory of resultants needed for the discussions that follow.  In Section \ref{sec:Factorisation} we outline how the use of iterated resultants in CAD leads to factorisations and then in Section \ref{sec:discard} we describe how in the case of multiple equational constraints in the input we could make savings through the use of multivariate resultants.  These three sections reproduce our SC-Square 2023 Workshop paper \cite{DavenportEngland2023a} with important additions to give clarity to the suggested optimisation and the impact it would have, illustrated by a new example and complexity analysis.

After this, in Section \ref{sec:MB}, we revisit \cite{McCallumBrown2009} which described an alternative approach to make such savings in the case of just two equational constraints.  We correct a typo in the original paper \cite{McCallumBrown2009}, provide a missing proof from that paper, and discuss possible extensions, thereby laying the foundation for future work in this area. While this is, strictly speaking, a separate kind of approach with respect to the one described in Section \ref{sec:discard},
it does fit within the general framework of exploring the use of multivariate resultants and discriminants in developing CAD-like algorithms, for use when appropriate, having less reliance on iterated resultant and discriminant computation.

The original motivation and a common application of CAD is Real QE (for how CAD may be utilised for this we again refer to the collection \cite{CavinessJohnson1998}).  The theory of both CAD and Real QE is usually expressed in polynomials, however in practice, we are often faced with problems defined using rational functions.  In \cite{Uncuetal2023a} we proposed two ways that this could be addressed in the case of SMT for non-linear real arithmetic, i.e. the case where formulae are fully existentially quantified.  In Section \ref{sec:denominators} we extend this to the more general case of Real QE, i.e. for any quantifier structure.  We start by noting, via an example, how the straightforward approach of imposing that the denominator be non-vanishing is not sufficient for universal quantifiers, before describing a better approach.

\section{Resultants}
\label{sec:resultants}

\subsection{Univariate resultants}
\label{subsec:unires}

Consider two polynomials $f,g\in K[y,x] = K[y_1,\ldots,y_s,x]$, where $K$ is a field, regarded as polynomials in $x$ whose coefficients depend on the $y_j$. 
The (univariate) resultant of these two polynomials is another polynomial $r(y) \in K[y]$ formed of their coefficients that is equal to zero if and only if $f$ and $g$ have a common root in an extension of $K$ or their leading coefficients $a(y), b(y)$ both vanish:
\[
r(\beta)=0 
\Leftrightarrow 
[(\exists x \in \bar{K} f(\beta, x)=g(\beta, x)=0) 
\vee (a(\beta) = b(\beta) = 0)],
\]
where $\beta = (\beta_1,\ldots,\beta_s) \in K^s$, and $\bar{K}$ denotes the algebraic closure of $K$. Resultants are a widely used tool in symbolic computation, and also in satisfiability checking over non-linear real arithmetic.  In particular, they are a key ingredient of CAD which traditionally requires many iterated univariate resultant calculations during projection \cite{Collins1975}. Basic algebraic properties of resultants are found in algebra texts such as\cite{CoxLittleOShea1997}; for computational aspects see \cite{Collins1971}.

\subsection{Notation}
\label{subsec:notation}

As usual, we will use $\res$ (or $\res_z$ when we need to make the variable explicit) to denote the resultant, and $\disc$ (or $\disc_z$) to denote the discriminant, the main tools used in CAD projection.  

The univariate resultant is often defined constructively in relation to the determinant of the Sylvester matrix of the two polynomials.  However, for use in CAD we implicitly work with a square-free basis (and often an irreducible basis to avoid having to make retrospective changes when new polynomials are added to the mix\footnote{One further advantage of an irreducible basis is described under $C$ in \S\ref{subsec:Issue}.}).  Thus for use in CAD we ignore repeated or common factors as in the following example.

\begin{ex}
Let $f=x^2-1$ and $g=x^2-2x+1$ which, since they have a common root, means their resultant using the traditional definition is zero.  

However, for CAD we interpret their resultant as $\res_x\left((x-1)(x+1),(x-1)^2\right)$ and then since we delete common factors this becomes $\res_x\left((x+1),(x-1)\right)$, which evaluates to 2.
\end{ex}

\begin{rem}\label{convention1}
When we use the notation $\res$ or $\disc$ in this paper we will always assume that common factors have been removed from the input, so as not to make the resultant zero.
Repeated factors can also be removed, since $\res(f,g^2)=\res(f,g)^2$ etc., and we will also assume this has also been done.
\end{rem}

\subsection{Multivariate resultants}
\label{subsec:mresTheory}

We are grateful to \cite{McCallum1999b, McCallumWinkler2018a} for clear expositions of certain results from \cite{Jouanolou1991}, which we have borrowed. The setting is polynomials in $x_1,\ldots,x_n$ with coefficients which are either indeterminates (of which we can take integer combinations as in Definition \ref{def:inertia}) or elements of $K[y]$, where $K$ is a field and $y$ represents an $s$-tuple of different indeterminates.

\begin{defn}\label{def:inertia}
Given $r$ homogeneous polynomials $F_1, \ldots, F_r$ in $x_1, \ldots, x_n$, with indeterminate coefficients comprising a set $A$, an integral polynomial $T$ in these indeterminates (that is, $T \in \Z[A]$) is called an \emph{inertia form} for $F_1,{\ldots}, F_r$ if $x_i^\tau T \in (F_1,{\ldots}, F_r)$, the ideal generated by the $F_i$s, for suitable $i$ and $\tau$.
\end{defn}

Van der Waerden \cite[\S 81]{vanderWaerden1950}\footnote{A less explicit version can be found in \cite[\S 16.5]{vanderWaerden1970}.} observes that the inertia forms comprise an ideal $I$ of $\Z[A]$, and he shows further that $I$ is a prime ideal of this ring. It follows from these observations that we may take the ideal $I$ of inertia forms to be the resultant system for the given $F_1,{\ldots}, F_r$ in the sense that for special values of the coefficients from field $K$, the vanishing of all elements of the resultant system is necessary and sufficient for there to exist a non-trivial solution to the system $F_1 = 0,{\ldots}, F_r = 0$ in some extension of $K$.

Now consider the case in which we have $n$ homogeneous polynomials in the same number $n$ of variables \cite[\S 82]{vanderWaerden1950}. Let $F_1,{\ldots}, F_n$ be $n$ generic homogeneous polynomials in $x_1,{\ldots}, x_n$ of positive total degrees $d_1,{\ldots}, d_n$. That is, every possible coefficient of each $F_i$ is a distinct indeterminate, and the set of all such indeterminate coefficients is denoted by $A$. Let $I$ denote the ideal of inertia forms for $F_1,{\ldots}, F_n$. Proofs of the following two propositions may be found in \cite{vanderWaerden1950, McCallumWinkler2018b}.

\begin{prop}\cite[Proposition 5]{McCallumWinkler2018a} 
$I$ is a nonzero principal ideal of $\Z[A]$: $I = (R)$, for some $R \ne 0$.
$R$ is uniquely determined up to sign. We call $R$ the (generic multivariate) resultant of $F_1,{\ldots}, F_n$; and we may write
$R = \res(F_1, \ldots, F_n)$.
\end{prop}

\begin{prop}\cite[Proposition 6]{McCallumWinkler2018a}  
The vanishing of $R$ for particular $F_1,{\ldots}, F_n$ with coefficients in a field $K$ is
necessary and sufficient for the existence of a non-trivial zero of the system $F_1 = 0,{\ldots}, F_n = 0$ in some extension of $K$.
\end{prop}

The above considerations also lead us to the notion of a multivariate resultant of $n$ non-homogeneous polynomials in $n - 1$ variables $x_1, \ldots, x_{n-1}$ over $K[y]$, where $K$ is a field and $y = (y_1, \ldots, y_s)$ is a list of different indeterminates. Consider a given non-homogeneous polynomial $f(y, x_1,{\ldots}, x_{n-1})$ over $K[y]$ of total degree $d$ (in the $x_j$ indeterminates).  We may write $f = H_d + H_{d-1} + \cdots + H_0$, where the $H_j$ are homogeneous of degree $j$ (in the $x_j$). Then $H_d$ is known as the leading form of $f$ (with respect to the $x_j$). Recall that the homogenization $F(y, x_1,{\ldots}, x_n)$ of $f$ is defined by $F = H_d + H_{d-1} x_n + \cdots + H_0 x_n^{d_n}$. Let $f_1,{\ldots}, f_n$ be particular non-homogeneous polynomials in $x_1,{\ldots}, x_{n-1}$ over $K[y]$ of positive total degrees $d_i$ (in the $x_j$), and with leading forms $H_{i,d_i}$. We set $\res_{x_1, \ldots, x_{n-1}}(f_1,{\ldots}, f_n) = \res(F_1,{\ldots}, F_n)$ , where $F_i$ is the homogenization of $f_i$. The following proposition displays a fundamental property of this multivariate resultant (see proofs in \cite{McCallum1999b, McCallumWinkler2018b}).

\begin{prop}\cite[Theorem 2.4]{McCallum1999b}
Let $\bar{K}$ denote the algebraic closure of $K$.
The vanishing of $\res_{x_1, \ldots, x_{n-1}}(f_1,{\ldots}, f_n)$ for a particular value $y = \beta$ in $\bar{K}^s$ is necessary and sufficient for 
\begin{itemize}
\item[(a)] {\bf either} the forms $H_{i,d_i}(\beta, x_1, \ldots, x_{n-1})$ to have a common nontrivial zero over $\bar{K}$,
\item[(b)] {\bf or} the polynomials $f_i(\beta, x_1, \ldots, x_{n-1})$ to have a common zero over $\bar{K}$.
\end{itemize}
\end{prop}

The above proposition is a generalization of Theorem 5 of \cite{Collins1971},
which is basic to the theory of the univariate (that is, Sylvester) resultant, and was briefly recalled in Section \ref{subsec:unires}.
For geometric intuition about the above proposition, observe that in the first case, non-trivial common zeros of the leading forms correspond to the projective solutions on the hyperplane at infinity; 
whereas in the second case common zeros of the $f_i$ correspond to the affine solutions of the system.

We note that practical aspects of computing multivariate resultants are discussed in Chapter 3 (especially Section 4) of \cite{CoxLittleOShea1998}.

\section{Factorisation of Iterated Univariate Resultants}
\label{sec:Factorisation}

Collins observed and suggested in \cite[pp. 177--178]{Collins1975} that iterated resultants, where there are ``common ancestors'', tend to factor. This was responded to by van der Waerden in a personal letter to Collins \cite{vanderWaerden1975}, which comprised relatively intuitive justifications of the observed factorizations. The letter was studied in the 1990s by one member of our present author group, and its content (in part) was formalized in \cite{McCallum1999b} as a collection of generic factorization theorems. Further developments were reported in \cite{BuseMourrain2009,LazardMcCallum2009}.   The work in \cite{McCallum1999b} is based on the theory in \cite{vanderWaerden1950}, which \cite{Jouanolou1991} notes has been deleted from more recent editions (such as \cite{vanderWaerden1970}). The work in \cite{BuseMourrain2009} is based on that of \cite{Jouanolou1991}.  

Despite this factorisation being observed since the inception of CAD, and subsequently proved rigorously, we are not aware of any optimisations in CAD implementations in regards to it.

\subsection{Example}

Consider these polynomials:
\begin{align*}
f &= y^2 + z^2 + x + z - 1, \\
g &= -x^2 + y^2 + z^2 - 1, \\
h &= x^2 + y + z.
\end{align*}
Under variable ordering $z \succ y \succ x$ we may calculate the iterated resultant:
\begin{equation}\label{eq:2}
\begin{array}{rcl}
\res_y(\res_z(f,g),\res_z(f,h))&=&5 x^{8}+16 x^{7}+14 x^{6}-2 x^{5}-12 x^{4}-8 x^{3}+3 x^{2}+2 x\\
&=&\underbrace{x \left(5 x^{3}+6 x^{2}-3 x -2\right)}_{\hbox{spurious}}\underbrace{\left(x^{2}+x +1\right) \left(x^{2}+x -1\right)}_{\hbox{genuine}}
\end{array}.
\end{equation}
We define the meaning of the labels ``\emph{spurious}'' and ``\emph{genuine}'' below.  First note that an alternative computational path may have calculated similarly
\begin{equation}\label{eq:3}
\begin{array}{rcl}
\res_y(\res_z(f,g),\res_z(g,h))&=&5 x^{8}+16 x^{7}+18 x^{6}+8 x^{5}-5 x^{4}-8 x^{3}-2 x^{2}+1\\
&=&\underbrace{\left(x^{2}+x +1\right) \left(x^{2}+x -1\right)}_{\hbox{genuine}} \underbrace{ \left(5 x^{4}+6 x^{3}+x^{2}-1\right)}_{\hbox{spurious}}.
\end{array}
\end{equation}
The final choice would have been to calculate, 
\begin{equation}\label{eq:4}
\begin{array}{rcl}
\res_y(\res_z(f,h),\res_z(g,h))&=&2 x^{4}+4 x^{3}+2 x^{2}-2 
\\
&=& 2\underbrace{ \left(x^{2}+x +1\right) \left(x^{2}+x -1\right)}_{\hbox{genuine}}.
\end{array}
\end{equation}
Up to constants (\ref{eq:4}) divides (\ref{eq:3}) and (\ref{eq:2}), but this need not happen in general. What does happen in general is that, if we consider a Gr\"{o}bner Basis,
\begin{equation}
\verb+Basis+_{\tt plex}(f,g,h)=\left\{
\left(x^{2}+x +1\right) \left(x^{2}+x -1\right), 
y -x, 
x^{2}+x +z
\right\},
\end{equation}
then we see that the basis polynomial in $x$ only divides all three iterated resultants and in fact \emph{is} $\res_{y,z}(f,g,h)$ in the sense of Section \ref{subsec:mresTheory}. In this example, it is also (\ref{eq:4}), but again this need not happen in general.  We note that the example calculations in this and subsequent sections are done in Maple 2023: we provide the Maple worksheet and a pdf printout as Supplementary Material.

The labels above are made in regards to the roots of the tagged resultant factors.  For (\ref{eq:2}) and the two alternative computations, the roots of the part we have labelled as ``genuine'' are 
\begin{equation}\label{eq:root1}
\{x:\exists y\exists z f(x,y,z)=g(x,y,z)=h(x,y,z)=0\}.
\end{equation}
 For (\ref{eq:2}), the set of roots of the part we have labelled as ``spurious'' is the set difference of
\begin{equation}\label{eq:root2}
\left\{x :\exists y\left(\exists z_1 f(x,y,z_1)=g(x,y,z_1)=0\land \exists z_2\ne z_1 f(x,y,z_2)=h(x,y,z_2)=0\right)\right\}
\end{equation}
and (\ref{eq:root1}). 
There are slightly different definitions for the two alternatives.
They are ``spurious'' in the sense that they do not extend to form true simultaneous zeros of $f,g,h$.  Nevertheless, they are $x$ values above which the topology changes, so they cannot always be discarded. Note that Section \ref{subsec:mresTheory} implies that there is always a neat factorisation (over $\Z$ if that was the original ring) into ``genuine'' versus ``spurious'' factors that depend on the non-eliminated variables: factors that are in the content with respect to the non-eliminated variables are not considered here, as they do not take part in further projections.

\subsection{Effect of the variable ordering on the example}

What happens if we take the variables in a different order?  In ordering $x \succ y \succ z$ we have: 
\begin{equation}
\res_y(\res_x(f,g),\res_x(f,h))=(z^2-1)^2,
\end{equation}
\begin{equation}
\res_y(\res_x(f,g),\res_x(g,h))=(z^2-1)^4,
\end{equation}
\begin{equation}
\res_y(\res_x(h,g),\res_x(f,h))=(z^2-1)^4,
\end{equation}
and 
\begin{equation}
\verb+Basis+_{\tt plex(x,y,z)}(f,g,h)=\left\{z^2-1,y^2+y+z,x-y\right\}.
\end{equation}
I.e. no spurious roots were uncovered with this ordering.  The question of CAD variable ordering is well studied and known to greatly affect the complexity of CAD both in practice \cite{delRioEngland2022c} and in theory \cite{BrownDavenport2007}.  The introduction of spurious factors in some orderings but not others may be a significant contributing factor to this.  

\section{Discarding Spurious Factors in CAD}
\label{sec:discard}

\subsection{Equational constraints}
\label{subsec:ECs}

One situation where the spurious factors can be concluded redundant is the case of CAD with equational constraints.
\begin{defn}
An \emph{equational constraint} is an equation whose truth is logically implied by the input formula under study (either explicitly, appearing as a conjunction to the rest of the formula, or otherwise implicitly). 
\end{defn}
Intuitively, the presence of an equational constraint reduces the dimension of the possible solution set by one, and so we may hope to reduce some of the computation in finding that solution set accordingly.  There has been work to this effect for CAD: starting with \cite{McCallum1999a} and summarised  in \cite{Englandetal2019a}.  In particular, we utilise later \cite{McCallum2001} which defined a reduced CAD projection operator $P^{*}_{\mathcal{E}}(\mathcal{A})$ for calculating the CAD projection of a polynomial set $\mathcal{A}$ in the presence of an equational constraint whose factors are in $\mathcal{E} \subset \mathcal{A}$.  

Consider the case of $k$ (explicit) equational constraints, along with some additional constraints, all expressed in variables $x_1, \dots, x_n$:
\begin{equation}\label{eq:setting}
\Phi\equiv f_1=0\land f_2=0\land\cdots f_k=0 \land \overline\Phi(f_{k+1},\ldots,f_m).
\end{equation}
The theory developed in \cite{McCallum2001} allows us to project with $\res_{x_n}(f_1,f_i)$, $i > 1$,
and $\disc_{x_n}(f_i)$, $i \ge 1$, as well as various coefficients, which do not contribute to the degree explosion.  
In particular, we do not need any $\res(f_i, f_j)$ for $1 < i < j$ which would be included in a CAD projection without knowledge of equational constraints. 
Note that here $f_1$ was selected as the \emph{designated pivot} equational constraint for the projection 
(but any other equational constraint would have been valid to designate as pivot). We may denote our first projection set by $J_{n-1}$. 

After the first projection we cannot use any $f_i$ as an equational constraint for the next projection, 
however, we can still make savings through the presence of implicit equational constraints.  
E.g., since $f_1=0$ and $f_2=0$, we know that $\res_{x_n}(f_1,f_2)=0$ also. 
(In fact, all the $\res_{x_n}(f_1,f_i)$ are equational constraints.) Let us select $\res_{x_n}(f_1, f_2)$ as the pivot constraint at level $n-1$,
and denote this polynomial by $g_1(x_1, \ldots, x_{n-1})$.
Then for the next projection we may take:
\begin{equation}\label{eq:resres}
\res_{x_{n-1}}(g_1,\res_{x_n}(f_1,f_i)), i > 2,
\end{equation} 
$\res_{x_{n-1}}(g_1,\disc_{x_n}(f_i))$, $i \ge 1$,
$\res_{x_{n-1}}(g_1, c)$, for all $(n-1)$-variate coefficients $c \in J_{n-1}$,
and discriminants and (enough) coefficients of all elements of $J_{n-1}$.
Note that we have presented here a slightly simplified description of the first two steps of the equational projection scheme by omitting to take explicit account of squarefree (or irreducible) basis computation at level $n-1$; but such basis computation must be performed in practice.

\subsection{Projection complexity with equational constraints}
\label{subsec:complexity1}

Traditional CAD projection to produce a sign-invariant decomposition for a set of polynomials was shown to produce doubly exponentially many polynomials who each have doubly exponential degree (in both cases taken with respect to the number of projections).   The projection operator from \cite{McCallum2001} reduces this by taking advantage of equational constraints (producing a decomposition truth invariant for the logical formula).  See \cite{Englandetal2019a} for a full complexity analysis of the various savings available from equational constraints to date.  To summarise, we can extract savings sufficient to reduce the double exponent in the number of polynomials, but not their degree.  

We reproduce some of this analysis to demonstrate the issue we try to address in the next subsection.  We will observe the growth in projection polynomials, focussing on resultants and discriminants (ignoring coefficients) and it is these which exhibit degree growth.

Consider (\ref{eq:setting}) and assume each $f_i$ is of degree at most $d$ (degree taken in each variable separately).  Hence we start with $m$ polynomials of degree $d$.   Using the projection operator from \cite{McCallum2001}, after the first elimination, we have 
\begin{equation}\label{eq:setting2}
\underbrace{\res_{x_n}(f_1,f_2), \ldots \res_{x_n}(f_1,f_m)}_{(m-1) \times \hbox{ degree }2d^2}, \underbrace{\disc_{x_n}(f_1),\ldots\disc_{x_n}(f_m)}_{m \times \hbox{ degree }2d(d-1)}.
\end{equation}
For the sake of simplicity, regard these as $2m-1$ polynomials $g_i$ of degree at most $2d^2$.  Selecting $g_1=\res_{x_n}(f_1,f_2)$ as the next designated pivot equational constraint, we have the following after the second elimination:
\begin{equation}\label{eq:setting3}
\underbrace{\res_{x_{n-1}}(g_1,g_2), \ldots \res_{x_{n-1}}(g_1,g_{2m-1})}_{(2m-2) \times \hbox{ degree }8d^4}, \underbrace{\disc_{x_{n-1}}(g_1),\ldots\disc_{x_{n-1}}(g_{2m-1})}_{(2m-1) \times \hbox{ degree }8(d^4-d^2)}.
\end{equation}
For the sake of simplicity, regard these as $4m-3$ polynomials $h_i$ of degree at most $8d^4$.  If we follow  \cite{McCallum2001} then we must select one $h_i$ as designated pivot equational constraint for the third elimination, say $h_1=\res_{x_{n-1}}(g_1,g_2)=\res_{x_{n-1}}(\res_{x_n}(f_1,f_2),\res_{x_{n}}(f_1,f_3))$, and obtain: 
\begin{equation}\label{eq:setting4}
\underbrace{\res_{x_{n-2}}(h_1,h_2), \ldots \res_{x_{n-2}}(h_1,h_{4m-3})}_{(4m-4) \times \hbox{ degree }128d^8}, \underbrace{\disc_{x_{n-2}}(h_1),\ldots\disc_{x_{n-2}}(h_{4m-3})}_{(4m-3) \times \hbox{ degree }128(d^8-d^4)}.
\end{equation}
The equivalent of (\ref{eq:resres}) after $k$ eliminations (i.e. eliminating all equational constraints) has degree $\Oo\big((2d)d^{2^k}\big)$ (i.e. doubly exponential).

\subsection{Suggested use of multivariate resultants}
\label{subsec:simpleMres}

In the setting of (\ref{eq:setting}), we are really only interested in the genuine zeros of the equational constraints, since away from these the formula will be uniformly false and thus further refinement is unnecessary.  Our suggestion is thus to use utilise multivariate resultants for the designated pivot equational constraints in such cases.

The first and second eliminations will proceed as above, but we can deviate at the third and subsequent eliminations.  As described above, for the third elimination we would usually designate as pivot equational constraint some
\[
A:=\res_{x_{n-1}}(\res_{x_n}(f_1,f_2),\res_{x_n}(f_1,f_i)) 
\]
for $i>2$ (for simplicity we suggest to take $i=3$).  However, this is ``\emph{overkill}'', as it includes the spurious roots defined as in Equation \eqref{eq:root2},  as well as the genuine simultaneous roots of all three polynomials.  Instead, let us use
\[
B:=\res_{x_{n-1}, x_n}(f_1,f_2,f_i)
\]
(again, with the suggestion to take $i = 3$)
as the designated pivot equational constraint, which will in general be much shorter.

We note that in the earlier workshop version of this paper \cite{DavenportEngland2023a} we made this suggestion with somewhat sloppy language.  Most importantly\footnote{The other omissions in the corresponding statement in \cite{DavenportEngland2023a} concerned the restrictions on $i$ and the variables to eliminate!}, we did not clarify that we must still leave all the factors of $A$ within the set of projection polynomials found after the second elimination, i.e. $\mathcal{J}_{n-2}$ using the notation from \cite{McCallum2001}.  The suggestion is simply to take the irreducible factors of $B$ (which are themselves amongst the irreducible factors of $A$) to form $\mathcal{E}_{n-2}$ in the computation of $P^{*}_{\mathcal{E}_{n-2}}(\mathcal{J}_{n-2})$.

We emphasise that we are not changing the set of projection factors, just which factors are used to form the designated pivot equational constraint. The correctness of this approach largely rests on the simple fact that $B$ is itself an equational constraint of the input, albeit an implicit one, opening the door to the use of the existing theory of \cite{McCallum2001}. 

A careful reader may notice that the theoretical framework of \cite{McCallum2001} is not quite adequate for designating $B$ as a pivot equational constraint at level $n-2$.  The reason is that \cite{McCallum2001} implicitly assumes that any candidate pivot constraint $e_k$ at level $k$ ($1 < k < n$) is a product of elements of $\mathcal{J}_k$, the set of level $k$ projection polynomials.  This was to ensure $\mathcal{E}_k \subseteq \mathcal{J}_k$, and hence that the subsequent semi-restricted projection was genuinely contained in the unrestricted one.  This appears a problem for us as the multivariate resultant $B$ need not be a product of elements of $\mathcal{J}_{n-2}$.  Fortunately, we may easily propose a remedy.  We designate $e_{n-2} = B$ and $\mathcal{E}_{n-2} = \{e_{n-2}\}$, and observe $\mathcal{F}_{n-2} \subseteq \mathcal{B}_{n-2}$, where $\mathcal{F}_{n-2}$ and $\mathcal{B}_{n-2}$ are finest squarefree bases for $\mathcal{E}_{n-2}$ and $\mathcal{J}_{n-2}$, respectively (by \cite{McCallum1999b}).  This ensures that the subsequent semi-restricted projection is genuinely contained in the unrestricted one.

\begin{rem}
The theory of \cite{McCallum1999b,McCallum2001} is cast in terms of irreducible bases, partly for simplicity and partly to align with the well-established practice (used in QEPCAD, for example) of computing an irreducible basis at each level, following projection, on efficiency grounds. Hence to use this theory, we also need to speak in terms of irreducible bases, as in the previous paragraph. In terms of implementation, we can use any square-free basis, because we have justified that its irreducible factors are precisely what we need.
\par
Since factorisation is generally good for efficiency (compare computing $\res(f_1f_2,g)$ with separately computing $\res(f_1,g)$ and $\res(f_2,g)$), we can see that the primitive really needed is ``half-hearted factorisation'': do a square-free decomposition, and as much more as you can easily do, which would probably mean giving up at the recombination stage if the $p$-adic factors didn't correspond to true factors over the integers.
\end{rem}

The theory of \cite{McCallum2001} moreover requires that $B$ is order invariant 
but this will not require any changes to the projection process: 
the discriminant of $B$ will be produced already as it is a factor of the discriminant of $A$, 
and similarly for the required coefficient of the projection factors. 
With $B$ the designated pivot equational constraint we may query whether we need to add the resultant of $B$ with $A$: 
but note that the intersection of $B$ with those other factors of $A$ will already be produced 
as part of the discriminant of $A$ and so this would obtain no additional projection factors.

Thus, our suggestion is valid using the theory of \cite{McCallum2001} and does not produce any additional projection factors.  However, as we will see next, we should get savings by designating this smaller $B$ as equational constraint instead of $A$.

\subsection{Worked example}
\label{subsec:example}

Consider the polynomials $$\begin{array}{rcl}f_1&:=&\left(z -y \right)^{3}+\left(x -w \right)^{3}-\left(x -1\right) y +z w,\\f_2&:=& x^{3}+y^{3}+z \left(y -1\right)+w x,\\f_3&:=& w^{3}+z^{3}+w y +z x,\\f_4&:=&\left(z +x \right)^{3}+\left(w +y \right)^{3}-z x +w y,
\end{array}$$
where at least $f_1$, $f_2$ and $f_3$ are equational constraints. 

In what follows we do not write out full polynomials for brevity but give the number of terms in their expanded form and their total degree, $D$.  The corresponding calculations may be found in the Supplementary Material.  The multivariate resultant was calculated using the third party DR Package of Manfred Minimair \cite{Minimair2017}.
\begin{enumerate}

\item Select $f_1$ as the designated pivot equational constraint, and project out $w$ following \cite{McCallum2001}. The polynomials are then $\res_w(f_1,f_i)$ and $\disc_w(f_i)$.
Call these polynomials $g_1,\ldots,g_7$.

\item As $f_1,f_2$ are equational constraints, so is $g_1=\res_w(f_1,f_2)= \underbrace{-x^{9}-3 x^{6} y^{3}-3 x^{3} y^{6}-\cdots
}_{\hbox{40 terms, $D=9$}}
$.
We can select this as the designated pivot equational constraint and project out $z$, again following \cite{McCallum2001}. The polynomials are then $\res_z(g_1,g_i)$ and $\disc_z(g_i)$.
Call these polynomials $h_1,\ldots,h_{13}$.

\item $h_1=\res_z(g_1,g_2)=\res_z(\res_w(f_1,f_2),\res_w(f_1,f_3))$ is also an equational constraint. This is $\underbrace{-512 x^{81}-13824 x^{78}y^3+\cdots}_{\hbox{3186 terms, $D=81$}}$.

\noindent Since $f_1,f_2,f_3$ are all equational constraints, we could select $h_1$ as the designated pivot equational constraint, which is equivalent to $A$ in \hbox{(12) in \cite{DavenportEngland2023a}}. 

\noindent However, following the suggestion we can instead use the multivariate resultant 
\[
\widehat h_1=\res_{w,z}(f_1,f_2,f_3)=\underbrace{8 x^{27}+72 x^{24} y^{3}+288 x^{21} y^{6}+\cdots}_{\hbox{359 terms, $D=27$}}
\]
as the designated pivot equational constraint.

\item $h_2=\res_z(g_1,g_3)=\res_z(\res_w(f_1,f_2),\res_w(f_1,f_4))$ is the way $f_4$ appears when it comes to projecting out $y$.  In fact $h_2=\underbrace{512 x^{81}+13824 x^{78}y^3+\cdots}_{\hbox{3214 terms, $D=81$}}$. The options are
$$
\begin{array}{rcl}
\res_y(h_1,h_2)&=&\underbrace{-4.97\cdot10^{1265}x^{3894}+\cdots}_{\hbox{3651 terms, $D=3894$}},\\
\res_y(\widehat h_1,h_2)&=&\underbrace{-2.6\cdot10^{428}x^{1292}+\cdots}_{\hbox{1211 terms, $D=1292$}},
\end{array}
$$
which already shows substantial advantages. If the $f_i$ were denser, the savings in the number of terms would be greater still.

\item In contrast to both the above,
$$
\begin{array}{rcl}
\res_{w,z,y}(f_1,f_2,f_3,f_4)&=&\underbrace{-2.56\cdot10^{35}x^{70}+\cdots}_{\hbox{62 terms, $D=70$}},\\
\end{array}
$$
may be used if $f_4$ were an equation constraint and we were doing further projections.
\end{enumerate}

\subsection{Why keep the iterated resultant in the projection set?}
\label{subsec:Issue}

Why does $A$ have to stay in the projection set, $P^*_{\res_{x_n}(f_1,f_2)}(P_{f_1}^*(F))$, rather than being replaced by $B$ there also?  Write $A=BCD$, where $D$ is the product of those factors of $A$ relatively prime to $B$, and $C=A/(BD)$, i.e. the product of factors of $B$ which exist to higher multiplicity in $A$ than in $B$. So the question is: ``Why can we not ignore $C$ and $D$?''
\begin{description}
\item[$\bm{C}$] Let $b$ be a factor of a square-free decomposition of $B$. Then $B$ will be order-invariant on the zeros of $b$.  However, $A$ might not be, if $B$ factors further as $b_1b_2$, and these divide $A$ to different multiplicities.  This problem is easy to solve, and indeed goes away if we take an irreducible basis rather than just a square-free basis.
\item[$\bm{D}$] The roots of $D$ should correspond to the spurious roots, and thus $f_1$, $f_2$ and $f_i$ do not actually intersect here. This is true, but the assertions in \cite{McCallum2001} are not about the triple intersections, even though these are all we care about, so we would have to adjust the theory developed in \cite{McCallum2001} and reprove validity.  A further potential problem is that, although $D$ and $B$ are relatively prime, their varieties might still intersect, giving rise to a change in order of $A$, but not of $B$; a potential barrier to proving such validity given that the correctness of \cite{McCallum2001} is based on a theory of order-invariance.
\end{description}
In the case of only two equational constraints, a suitable change to the statements from \cite{McCallum2001} to let us lift directly to a variety defined by two constraints is given in \cite{McCallumBrown2009}.  We will revisit that in Section \ref{sec:MB}.

\subsection{Impact on degree growth}
\label{subsec:complexity2}

Does the suggestion of Section \ref{subsec:simpleMres} reduce the degree growth complexity outlines in Section \ref{subsec:complexity1}?.  We saw there that the iterated resultants after $k$ eliminations have degree $\Oo\big((2d)d^{2^k}\big)$.  In comparison, the theoretical multivariate resultant $\res(f_1,\ldots,f_k)$ has degree $\Oo\left(d^k\right)$ (the B\'{e}zout bound).  More precisely, the Dixon multivariate resultant is known to have degree $k!d^k$ (see \cite[\S 4]{KapurSaxena1995}).

Suppose we instead selected the multivariate resultant $h:=\res_{x_n,x_{n-1}}(f_1,f_2,f_3)$ as designated pivot equational constraint for the third elimination.  Then instead of (\ref{eq:setting4}) we get 
\begin{equation}\label{eq:setting5}
\underbrace{\res_{x_{n-2}}(h,h_2), \ldots \res_{x_{n-2}}(h,h_{4m-3})}_{(4m-4) \times \hbox{ degree }96d^7}, \underbrace{\disc_{x_{n-2}}(h_1),\ldots\disc_{x_{n-2}}(h_{4m-3})}_{(4m-3) \times \hbox{ degree }128(d^8-d^4)}.
\end{equation}
Note that while as discussed in Section (\ref{subsec:simpleMres}), we take the discriminant of the larger $h_1$ rather than that of the smaller $h$.   

At this stage, the saving does not seem great: $96d^7$ instead of $128d^8$.  This is because no savings were available for the first and second elimination, with only the third allowing a smaller designated pivot equational constraint.  However, from now on all projections (until $k$ assuming the $k$ equational constraints were independent) can make use of a multivariate resultant.  So for the fourth elimination we are comparing the use of an equational constraint of degree $\Oo(d^8)$ if we follow \cite{McCallum2001}, to a multivariate resultant of degree $\Oo(d^4)$ in the form of $\res_{x_n,x_{n-1},x_{n-2}}(f_1,f_2,f_3,f_4)$ using this methodology of Section \ref{subsec:simpleMres} (not one of the $\Oo(d^7)$ resultants in (\ref{eq:setting5})).  

In summary, while \cite{Englandetal2015a} observed that the use of existing optimisations for $k$ equational constraints reduces the double exponent of $m$ (the number of polynomial constraints) from $n$ to $n-k$; here we show that the suggestion in Section \ref{subsec:simpleMres} gives the same reduction in the double exponent of $d$ (the degree of the constraints) \emph{in regards to the nested resultants}.  However, the overall degree growth is not controlled due to the discriminants in use.  In Section \ref{sec:MB} we consider one alternative that might be pursued in the future to counter this.

\subsection{Detecting spurious factors}
\label{subsec:detecting}

In the examples above the factors were marked as ``spurious'' or ``genuine'' via  manual analysis to see if the roots of the factors led to common zeros or not.  Are there alternatives to such manual detection?

We note that in some cases we can discard factors based on their degree. when this reaches the B\'{e}zout bound on the true multivariate resultant.  I.e., if $\res_y(\res_z(f,g),\res_z(f,h))$ has an irreducible factor of degree $>d^3$, it \emph{must} be spurious and can be discarded. Since it is common for CAD implementation to factor polynomials, this is a cheap, if incomplete, test.

\begin{ex}
For example, the following three 3-variable polynomials were created randomly in Maple to have total degree 5.  The computations described may be found in the Supplementary Material.
\begin{align*}
f &= -34x^2z^3 - 20y^5 + 7x^2y^2 - 43y^3z + 63x + 16z, \\
g &= 13xz^4 - 27z^4 - 21xy^2 + 30yz - 42x - 81, \\
h &= -65xz^4 + 13z^5 + 30x^3z + 17xy^3 + 25yz + 78.
\end{align*}
Then $\res_y( \res_z(f,g), \res_z(f,h) )$ factors into a constant times two irreducible polynomials:  one of degree $378$ and the other of degree $89$.  With no further computation we can identify the first as spurious since its degree is greater than $5^3 = 125$.  The second could be genuine, or be another spurious factor:  we may check manually that it is indeed genuine.
\end{ex}

In an example where we have multiple factors below the bound we could work through them in turn keeping count of the sum of degrees of genuine factors as we uncover then, in each case reducing the degree bound accordingly for any further factors to be investigated as genuine.

\subsection{Alternative projection operators}

Though it would have to be proved, it seems very likely that the suggestions and conclusions of this section, which were all developed in the McCallum projection theory \cite{McCallum1998, McCallum1999a, McCallum2001, Englandetal2015a},  would apply also to equational constraints within the Lazard projection theory \cite{McCallumetal2019a, Davenportetal2023a}. Here, equational constraints introduce fresh challenges with ``curtains'' \cite{Nair2021b}, similar to the challenges of nullification in \cite{McCallum1984}.

\section{Delineability of Varieties for Equational Constraints}
\label{sec:MB}

In Section \ref{sec:discard}, we suggested a fairly simple way to adapt CAD in the case of multiple equational constraints to avoid some of the spurious roots.  The optimisation suggested is easy to incorporate into existing implementations, and does not require the development of new theory.  However, the savings are essentially limited, as outlined in Section \ref{subsec:Issue}, to the projection phase rather than the lifting phase.  

An alternative approach was outlined in \cite{McCallumBrown2009} by one of the present authors for the  case of two equational constraints.  Here the authors defined $V \subset \R^r$ to be the real algebraic variety defined by the conjunction of the two equational constraints $f = 0 \land g = 0$, whose defining polynomials are in $x_1, \dots x_r$.  They then defined a new notion of the variety $V$ being delineable on a cell in $\R^{r-2}$, with respect to the last two variables $x_{r-1}$ and $x_r$; 
in contrast to traditional CAD where delineability is defined for a single $r$-variate polynomial $f$ on a cell in $\R^{r-1}$, with respect to the last variable $x_r$. This allowed for the idea of projection and lifting between $\R^r$ and $\R^{r-2}$ each being a single step.   

Along with defining this new delineability concept \cite{McCallumBrown2009} presented two main theorems.  The first, Theorem 4.1, described conditions that allowed us to conclude delineability of the variety over the cell. The second, Theorem 4.2, aimed to describe conditions for other polynomials present to be sign invariant upon the cells of the variety.  Together, these were to validate a new projection operation to use in this case, making use of so-called generalised resultants and discriminants.

This offers a promising avenue for future development.  In this paper we provide some correction and clarification on those first ideas in \cite{McCallumBrown2009}. We then formulate and discuss a conjecture indicating the kind of extension we could aim for.

\subsection{Corrected statement of Theorem 4.2 from (McCallum and Brown, 2009)}

The theorems used the notation that $x$, $y$ and $z$ denote the $(r - 2)$-tuple $(x_1,\ldots, x_{r- 2})$, and the variables $x_{r- 1}$, $x_r$, respectively, with $r \ge 3$. We recall for the reader's convenience the meaning of key terminology used in stating the theorems. Let $f$, $g$ and $h$ be $r$-variate real polynomials of positive degrees with respect to $y$ and $z$. Letting $\mathcal{R}$ denote $\R[x]$ and $I$ denote the ideal $(f, g, f_y g_z - g_y f_z)$ of $\mathcal{R}[y,z]$ (where $f_y$, $g_z$, etc. denote
partial derivatives), we call any element of the second elimination ideal $J = I \cap \mathcal{R}$ of $\mathcal{R}$ a {\em generalized discriminant} of $f$ and $g$ with respect to $y$ and $z$;
and we call any element of the second elimination ideal $(f, g, h) \cap \mathcal{R}$ of $\mathcal{R}$
a {\em generalized resultant} of $f$, $g$ and $h$ with respect to $y$ and $z$.

We further say that the real variety $V$ of $f$ and $g$ is {\em analytic delineable} on a submanifold $S$ of $\R^{r-2}$ with respect to $y$ and $z$ provided that there exist some $k \ge 0$ analytic functions $\theta_j : S \rightarrow \R^2$ whose graphs are disjoint such that the portion of $V$ inside the cylinder $S \times \R^2$ is the union of the graphs of the $\theta_j$ (which graphs are called the {\em sections} of $V$ over $S$);
and there exist positive integers $m_1, \ldots, m_k$ such that for every fixed point $x \in S$ and every $j$, the intersection multiplicity of $f(x,y,z)$ and $g(x,y,z)$ (considered as polynomials in $y$ and $z$ alone) at the common zero $\theta_j(x)$ is $m_j$.
Sections 1 and 2 of \cite{McCallumBrown2009} contain examples of the concepts just reviewed.

The first main theorem stated and proved in \cite{McCallumBrown2009} was as follows.
\def\foo{\cite[Theorem 4.1]{McCallumBrown2009}}
\begin{thm}[\foo]
Let $f(x,y,z)$ and $g(x,y,z)$ be relatively prime real polynomials
of positive degrees with respect to $y$ and $z$.
Let $D(x)$ be a generalized discriminant of $f$ and $g$ with respect to 
$y$ and $z$ and suppose that $D(x) \neq 0$.
Let $S$ be a simply connected submanifold of $\R^{r-2}$ such that the total number of common zeros, multiplicities counted, of $f$ and $g$ in $\C^2$
is finite and constant in $S$, and $D(x)$ is order-invariant in $S$.
Suppose further that there is no common zero of the polynomials
$f, g, f_z, f_y, g_z, g_y$ in the cylinder $S \times \R^2$ over $S$,
where $f_z, f_y$, etc. denote partial derivatives.
Then the variety $V$ of $f$ and $g$ is analytic delineable on $S$
with respect to $y$ and $z$.
\end{thm}

The second main theorem of \cite{McCallumBrown2009} was stated in the following way.
\def\foo{\cite[Theorem 4.2 with omission]{McCallumBrown2009}}
\begin{thm}[\foo]
Let
$f(x, y, z)$, $g(x, y, z)$ and $h(x, y, z)$ be real polynomials of positive
degrees with respect to $y$ and $z$. Let $R(x)$ be a generalised
resultant of $f$ and $g$ with respect to $y$ and $z$, and suppose
$R(x) \ne 0$. Let $S$ be a connected submanifold of $\R^{r- 2}$ on
which the variety $V$ of $f$ and $g$ is analytic delineable with respect to $y$ and $z$ and in which $R(x)$ is order-invariant. Suppose further that there is no common zero of the polynomials
$f$, $g$, $f_z$, $f_y$, $g_z$, $g_y$ in the cylinder $S\times\R^2$ over $S$. Then
$h$ is sign-invariant in each section of $V$ over $S$.
\end{thm}
The observant reader will notice that $h$ appears nowhere in the hypotheses, other than the statement of its existence. This makes it unlikely that we could draw conclusions about it. Indeed, the theorem requires a simple correction (insertion), which is made below in bold and red.
\def\foo{\cite[Theorem 4.2 corrected]{McCallumBrown2009}}
\begin{thm}[\foo]
Let
$f(x, y, z)$, $g(x, y, z)$ and $h(x, y, z)$ be real polynomials of positive
degrees with respect to $y$ and $z$. Let $R(x)$ be a generalised
resultant of $f$, $g$ {\color{red}\textbf{and} $\bm{h}$} with respect to $y$ and $z$, and suppose
$R(x) \ne 0$. Let $S$ be a connected submanifold of $\R^{r- 2}$ on
which the variety $V$ of $f$ and $g$ is analytic delineable with respect to $y$ and $z$ and in which $R(x)$ is order-invariant. Suppose further that there is no common zero of the polynomials
$f$, $g$, $f_z$, $f_y$, $g_z$, $g_y$ in the cylinder $S\times\R^2$ over $S$. Then
$h$ is sign-invariant in each section of $V$ over $S$.
\end{thm}

\subsection{Proof of corrected Theorem 4.2 from (McCallum and Brown, 2009)}

The proof of Theorem 4.2 was omitted in \cite{McCallumBrown2009} due to space constraints of the publication. For the sake of completeness, we give here the proof of (the corrected) Theorem 4.2.
\begin{proof}
Let $\sigma$ be a section of $V$, determined by the continuous (indeed, analytic) function
$\theta:S \rightarrow \mathbb{R}^2$, where $\theta(x) = (\theta_y(x), \theta_z(x))$ for all $x \in S$.
By connectedness of $\sigma$, it suffices to show that $h$ is sign-invariant in $\sigma$ near an arbitrary
point $(a,b,c)$ of $\sigma$.
That $h$ is sign-invariant in $\sigma$ near $(a,b,c)$ follows by continuity of $h$
in case $h(a,b,c) \neq 0$. So henceforth assume $h(a,b,c) = 0$.
By one of the hypotheses, at least one of $\partial f/\partial z$, $\partial f/\partial y$,
$\partial g/\partial z$, $\partial g/\partial y$ does not vanish at $(a,b,c)$.
Without loss of generality assume $\partial f/\partial z (a,b,c) \neq 0$ and $(a,b,c) = (0,0,0)$.

We aim to construct a function $R^*$, analytic near the origin in complex $(r-2)$-space $\mathbb{C}^{r-2}$,
whose zero set is the projection onto $\mathbb{C}^{r-2}$ of the portion of the
complex variety of $f$, $g$ and $h$ near the origin in $\mathbb{C}^r$.
This will require several steps, the first of which is as follows.
Since $\partial f/\partial z (0,0,0) \neq 0$, we may apply Hensel's lemma (Theorem 3.1 of \cite{McCallum1999b}) 
to $f$ near the origin in complex $(r-1)$-space $\mathbb{C}^{r-1}$.
By this lemma we conclude there is a polydisc $\Delta_1$ about
the origin and polynomials in $z$, $f_1(x,y,z) = z - \xi(x,y)$ and
$f_2(x,y,z)$, whose coefficients are elements of the formal power series ring $\mathbb{R}[[x,y]]$,
absolutely convergent in $\Delta_1$, such that $\xi(0,0) = 0$, $f_2(0,0,0) \neq 0$, and $f = f_1 f_2$.
Since a function defined as a sum of a convergent power series is analytic,
$\xi(x,y)$ and the coefficients of $f_2$ are analytic in $\Delta_1$.

The reader might guess that we could quickly complete the construction of the desired function $R^*$ by setting $R^*(x)$ to be a generalized resultant of $f_1$, $g$ and $h$ with respect to $y$ and $z$. However, since $f_1$ is not known to be polynomial in $y$ (though it is in $z$), such a proposed definition is not valid.  Consequently, a more involved process is needed.

For any $\delta > 0$, we denote by $\Delta(0; \delta)$ the disc in $\mathbb{C}$ about the origin of radius $\delta$.  The second step of our construction process follows. Put $P(x,y) = \res_z(f_1,g) = g(x,y,\xi(x,y))$, for all $(x,y) \in \Delta_1$. Where $m>0$ denotes the intersection multiplicity of $f(0,y,z)$ and $g(0,y,z)$ at $(0,0)$, we may infer that $y=0$ is a zero of $P(0,y)$ of multiplicity $m$ (by a slight extension of Theorem 5.3 of Chapter IV of \cite{Walker1962}, since the zeros of $f$ are the same as the zeros of $f_1$ near the origin, and the plane curves $f_1(0,y,z) = 0$ and $g(0,y,z) = 0$ have no intersections on the $z$-axis except at $z = 0$). Hence, by the $(r-1)$-variable analogue of the Weierstrass preparation theorem (as presented in Lecture 16 of \cite{Abhyankar1990}), there is a polydisc $\Delta_2 = \Delta_2' \times \Delta(0; \delta_2) \subset \Delta_1$ about the origin in $\mathbb{C}^{r-1}$, a polynomial $P_1(x,y) = y^m + a_1(x) y^{m-1} + \cdots + a_m(x)$, with the $a_i(x) \in \mathbb{R}[[x]]$, absolutely convergent in $\Delta_2'$, and an element $P_2(x,y)$ of $\mathbb{R}[[x,y]]$, absolutely convergent in $\Delta_2$, such that $P_1(0,y) = y^m$, $P_2(0,0) \neq 0$, and $P = P_1 P_2$.

To begin the third step of our construction, we put $T(x,y) = \res_z(f_1, h)$. 
By the Weierstrass division theorem (Lecture 16 of \cite{Abhyankar1990}), there exists a polydisc $\Delta_3 = \Delta_3' \times \Delta(0; \delta_3) \subset \Delta_2$, an element $Q(x,y)$ of $\mathbb{R}[[x,y]]$, absolutely convergent in $\Delta_3$, and an element $T^*(x,y)$ of $\mathbb{R}[[x]][y]$, of degree in $y$ at most $m-1$, whose coefficients are absolutely convergent in $\Delta_3'$, such that $T = P_1 Q + T^*$. By the analyticity (hence continuity) of the coefficients $a_i(x)$ of $P_1$, and polynomial root system continuity (Theorem (1,4) of \cite{Marden1966}), we may if needed refine $\Delta_3'$ to a smaller polydisc about the origin 	to ensure that for every fixed $x \in \Delta_3'$, each root of $P_1(x,y)$ is in $\Delta(0; \delta_3)$.

We may now complete our construction of $R^*$: we put $R^*(x) = \res_y(P_1, T^*)$. We claim that the zero set of $R^*$ is contained in the zero set of $R$ in $\Delta_3'$. The proof of this claim is as follows. Let $\alpha$ be an element of $\Delta_3'$ and suppose that $R^*(\alpha) = 0$. Then there exists $\beta \in \mathbb{C}$ such that $P_1(\alpha, \beta) = T^*(\alpha, \beta) = 0$. 	Since each root of $P_1(\alpha,y)$ is in $\Delta(0; \delta_3)$, we have $\beta \in \Delta(0; \delta_3)$.
Hence we can legally substitute $(\alpha, \beta)$ for $(x,y)$ in the power series identity $T = P_1 Q + T^*$, from which we deduce $T(\alpha, \beta) = 0$. The same substitution in the power series identity $P = P_1 P_2$ yields $P(\alpha, \beta) = 0$. Hence, with $\gamma = \xi(\alpha, \beta)$, we have 	$g(\alpha, \beta, \gamma) = h(\alpha, \beta, \gamma) = 0$. Substitution of $(\alpha, \beta, \gamma)$ into $f = f_1 f_2$ yields 	$f(\alpha, \beta, \gamma) = 0$. Therefore, by Theorem 2.3 of \cite{McCallumBrown2009} we have $R(\alpha) = 0$. The claim is proved.

Next, by Theorem 5.1 of \cite{BrownMcCallum2005}, there exists a polydisc $\Delta_4' \subset \Delta_3'$, an analytic function $R'$ in $\Delta_4'$, and an integer $n \ge 1$ such that $R^n = R^* R'$ in $\Delta_4'$. Since $R$ and $R^*$ have real power series representations, so does $R'$ (because the imaginary part of the power series expansion for $R'$ about the origin must be 0). 	Therefore, $R'$ is analytic in the box $B_4' = \Delta_4' \cap \mathbb{R}^{r-2}$. By Lemma A.3 of \cite{McCallum1988}, since $R$ is order-invariant in $S$ by hypothesis, 	$R^*$ is order-invariant in $S \cap B_4'$. But $R^*(0) = 0$, since $f_1$, $g$ and $h$ all vanish at the origin. Hence $R^*(x) = 0$, for all $x \in S \cap B_4'$.

We conclude our proof as follows. We claim that for every fixed $x \in S \cap B_4'$, $\theta_y(x)$ is a root of $P_1(x,y)$ of multiplicity $m$, hence the unique root of $P_1(x,y)$. To prove this claim, take some fixed $x \in S \cap B_4'$. Now the intersection multiplicity of the curves $f(x,y,z) = 0$ and $g(x,y,z) = 0$ at $(\theta_y(x), \theta_z(x))$ is $m$ since the intersection multiplicity of $f(0,y,z)$ 	and $g(0,y,z)$ at $(0,0)$ is $m$ (previously stated) and $V$ is analytic delineable with respect to $y$ and $z$ on $S$ (a key hypothesis). Therefore $y = \theta_y(x)$ is a root of $P(x,y)$, hence of $P_1(x,y)$, of multiplicity at least $m$, by a slight extension of Theorem 5.3 of Chapter IV of \cite{Walker1962}. Since $\deg_yP_1 = m$, the root $y = \theta_y(x)$ of $P_1(x,y)$ has multiplicity {\em exactly} equal to $m$, and hence this number is the unique root of $P_1(x,y)$. The claim is proved. 	

Our final claim is that $h(x, \theta_y(x), \theta_z(x)) = 0$, for all $x \in S \cap B_4'$. To prove this, take some fixed $x \in S \cap B_4'$. Then $R^*(x) = 0$ (established in the previous paragraph). Therefore, there exists $\beta \in \Delta(0; \delta_3)$ such that $P_1(x, \beta) = T^*(x, \beta) = 0$. But $y = \theta_y(x)$ is the unique root of $P_1(x,y)$ (already proved within this paragraph). Therefore $\beta = \theta_y(x)$. We deduce $T(x, \theta_y(x)) = 0$ (by substituting $(x, \theta_y(x))$ into the identity $T = P_1 Q + T^*$). Hence $h(x, \theta_y(x), \theta_z(x)) = 0$, by definition of $T$. 	Our final claim is proved, and the proof of the theorem is now complete.
\end{proof}

\subsection{Desired extension}

Let $\phi^*$ be a prenex formula in $r \ge 1$ variables of elementary real algebra of the form
$$
(Q_{f+1} x_{f+1}) \cdots (Q_r x_r) \phi(x_1, \ldots x_r),
$$
where the quantifiers are all existential or all universal.
We suppose that $\phi$ contains as a designated subformula
the conjunction of $t \ge 1$ equations 
$f_1 = 0 \wedge \cdots \wedge f_t = 0$,
where the $f_i$ are squarefree and pairwise relatively prime.
Furthermore, in case all quantifiers are existential,
we assume $\phi$ has the form 
$[f_1 = 0 \wedge \cdots \wedge f_t = 0] \wedge \psi$,
for some quantifier-free $\psi$; and in case all quantifiers
are universal, we assume $\phi$ has the form
$[f_1 = 0 \wedge \cdots \wedge f_t = 0] \implies \psi$,
for some quantifier-free $\psi$.

We would like to develop CAD-based methods, potentially more efficient
than the original general purpose CAD-based methods, for performing quantifier elimination
on such a formula $\phi^*$. Putting $p = r - f$ for notational convenience,
we note that the special cases $p = t = 1$ and $p = t = 2$ have been treated
in prior works. Indeed, the former case is included in \cite{McCallum1999a}, and the latter case is the content of \cite{McCallumBrown2009} discussed above.
So, it is natural to make some guesses about the general family of cases 
$p = t \ge 1$.

For each of the two studied cases $p = t = 1$ and $p = t = 2$
there are two key lifting theorems: one to ensure the delineability
of the real variety $V$ of the $f_i$ on a submanifold $S$ of $\R^f$
with respect to the last $p$ variables,
and the other to ensure that each of the remaining polynomials is sign-invariant in each section of $V$ over $S$.
For $p = t = 2$, the theorems from \cite{McCallumBrown2009} recalled (and, for the second lifting theorem, proved) in the previous subsection are these key results.

So, for the general case $p = t \ge 1$, we expect to require two
similarly worded lifting theorems. For the first such theorem,
we would need a notion of a generalized discriminant of the $t$
polynomials $f_1, \ldots, f_t$ with respect to the $t$ variables
$x_{f+1}, \ldots, x_r$. This could be defined by analogy with the definition for the case $t = 2$ described in Subsection 2.2 of \cite{McCallumBrown2009} and recalled in the previous subsection of this paper.
We would also need to define analogously the notion of the analytic delineability
of the real variety of the $f_i$ on a submanifold of $\R^f$, with respect to the last $t$ variables. The relevant conjecture is then:

\begin{con}
Let $f_1, \ldots, f_t$ be relatively prime real polynomials
of positive degrees with respect to $x_{f+1}, \ldots, x_r$.
Let $x$ denote $(x_1, \ldots, x_f)$.
Let $D(x)$ be a generalized discriminant of $f_1, \ldots, f_t$ with respect to $x_{f+1}, \ldots, x_r$ and suppose that $D(x) \neq 0$.
Let $S$ be a simply connected submanifold of $\R^{f}$ such that the total number of common zeros, multiplicities counted, of the $f_i$ in $\C^t$
is finite and constant in $S$, and $D(x)$ is order-invariant in $S$.
Suppose further that there is no common zero of the polynomials
$f_i$ and their first derivatives with respect to the last $t$ variables in the cylinder $S \times \R^t$ over $S$.
Then the variety $V$ of the $f_i$ is analytic delineable on $S$
with respect to the last $t$ variables.
\end{con}

We aim to prove this by suitably extending the proof of Theorem 4.1 from \cite{McCallumBrown2009}. The second general lifting theorem whose need is anticipated would then be obtained by straightforward generalization of Theorem 4.2 of \cite{McCallumBrown2009}.

\section{Rational Functions as Input}
\label{sec:denominators}

The theory of CAD is cast in terms of polynomials, but in practice, problems are often presented involving rational functions. How should we reconcile this? The SMTLIB format \cite{Barrettetal2021a} only allows polynomial constraints, and the same is true of, for example, the Tarski system \cite{ValeEnriquezBrown2018a}. 

In the ``New Perspectives in Symbolic Computation and Satisfiability Checking" Dagstuhl 2022 meeting \cite{Abrahametal2022a} allowing rational functions in the SMTLIB format and letting SMT-solvers handle them were discussed openly, following the talk of Duterte  \cite[\S 3.10]{Abrahametal2022a} . The main two concerns raised against allowing rational functions in SMTLIB language were ``What would happen if a solver re-packages and does arithmetic with an object of the form $x/0$?" and ``What if the solver splits the problem and, on a branch not directly related to the rational function, finds an assignment that makes the rational function undefined?" It was also made clear that defining objects of the form $x/0$, even when they respect equalities (i.e. $x=z \Leftrightarrow x/0 = z/0$), can be lead to proving wrong results. These concerns demotivated the cohort to work on the problem, let alone to come up with a unified solution. 

In the context of real satisfiability problems, these questions were addressed in an SC-Square 2022 Workshop paper by three of the present authors \cite{Uncuetal2023a} who introduced and compared two approaches, which we will recap shortly. Their approach was to interpret and preprocess rational functions in a mathematically consistent way to make the semantic concerns of the previous paragraph extraneous.  The formal manipulations suggested are easy to be carried out by SMT-solvers and this way rational functions can be allowed in SMT format as valid input.  While valid for the fully existential context of SMT, the discussion in \cite{Uncuetal2023a}  will be shown to be incomplete.  We introduce the problem and provide a proposed solution in this section for any  quantification structure.

\subsection{The problem with supposing non-zero denominators}
\label{subsec:CBProblem}

The material in this subsection was produced following an email we received from Chris Brown \cite{Brown2023f} stimulated in turn by a conversation he had with Zoltan Kov\'{a}cs: Zoltan wants to be able to send rational formulas to the Tarski software from his GeoGebra Discovery software \cite{Brownetal2021b}.  He suggested adding the requirement of non-vanishing denominators but Chris noted the problems with this by means of the following example.

Suppose the procedure is that, for a given atom $A$ which involves a non-trivial denominator $D$, we consider instead $\bar {A}\land D\not=0$, where $\bar{A}$ is some simplification of $A$ that is polynomial, e.g. multiplying up by the denominator $D$. 
Let us follow this procedure, looking at the problem 
\[
\forall x \left[\frac{1}{x^2} \geq 0 \right].
\] 
So the procedure lead us to study 
\[
\forall x \left[ {x^2}\not=0 \land {1} \geq 0\right].
\] 
For $x=0$, the statement on the line above is false (from the first clause). Consider next the negation of our original  example: $\exists x \, [1/{x^2}<0]$. The same procedure would lead us to study
\[ 
\exists x \left[ x^2\not = 0 \land 1<0 \right],
\]
which is also false (this time from the second clause). We cannot have a statement and its negation both false, which implies that the procedure was not a logical implication.

\subsection{Restricted domains and conditional universal quantification}
\label{subsec:restricted_domains}

The procedure above does not take into account that the original rational functions are defined on restricted domains.  The problem $\forall x [1/x^2 \geq 0]$ needs to be addressed within the domain of $1/x^2$, where all the points on which it is not defined (only $x=0$ in this case) should be left behind. More generally, we should be studying the following restricted quantification structure:
\[
\forall x, \underbrace{\text{for which }A\text{ is defined}}_{condition}, A.
\] 
However, most Quantifier Elimination procedures do not support such a statement directly.  The logical reduction of a conditional $\forall$ statement is an implication, which turns this problem to 
\[
\forall x \in \mathbb{R}, A \text{ is defined}\Rightarrow {A}.
\] 
Recall that logical implications make no assertion on what happens when the hypothesis is false.  I.e. one can replace the $``A\text{ is defined} \Rightarrow"$ directly with the conditions on the domain.  

\subsection{Prior work on denominators}

Rational functions and how to deal with the possible zeros of the denominators in the context of SMT for non-linear real arithmetic calculations have been a part of the SC-Square conversation, at least since the Dagstuhl 2022 meet \cite{Abrahametal2022a}. This led to the inclusion of Section 3 of our 2022 paper \cite{Uncuetal2023a} where we discussed how the problem could be handled in at least two ways, and further that the choice of approach led to different conclusions on the subsequent best SMT solver. 

Given polynomials $f(\vec x,y), g(\vec x, y) \in \mathbb{Z}[\vec x, y]$, with $f(\vec x,y)/g(\vec x,y)$ in minimal terms\footnote{i.e. there is no $(\vec x ,y)$ such that both $f(\vec x,y)=0$ and $g(\vec x,y)=0$ together.}, we suggested two possible equivalences:
\begin{align} 
\exists y \left[ \frac{f(\vec x,y)}{g(\vec x, y)} \, \sigma \, 0 \right] &\Longleftrightarrow \exists y [g(\vec x, y)\not =0 \land f(\vec x,y)g(\vec x, y) \, \sigma \, 0],\label{UDE-a}\\
\exists y  \left[ \frac{f(\vec x,y)}{g(\vec x, y)} \, \sigma \, 0 \right] 
&\Longleftrightarrow \exists y [(g(\vec x, y)>0 \land f(\vec x,y) \, \sigma \, 0) \label{UDE-b} \\
&\qquad \qquad \lor (g(\vec x, y)<0 \land 0 \, \sigma \, f(\vec x,y) )], \nonumber
\end{align}
where in both cases $\sigma$ is a relation operator from $\{=,\not=,>,<,\leq,\geq\}$. 

We argue now that these simplifications proposed in \cite{Uncuetal2023a} both take into account the implied conditions on the domain and thus do not suffer from the issues of the preceding sections.

Any problem involving only polynomials has the whole of $\mathbb{R}$ for each variable as its domain. However, while handling rational functions, under the hood, we have to deal with the domain of functions $f/g$. I.e., $\exists y  [{f(\vec x,y)}/{g(\vec x, y)} \, \sigma \, 0]$ does not translate to $\exists y \in \mathbb{R} [{f(\vec x,y)}/{g(\vec x, y)} \, \sigma \, 0]$, but instead translates to 
\[
\exists y \in\{z : g(\vec x,z)\not= 0)\} \left[ \frac{f(\vec x,y)}{g(\vec x, y)} \, \sigma \, 0 \right].
\]  
However, this is exactly the same as 
\[
\exists y \left[ g(\vec x,y)\not= 0\land \frac{f(\vec x,y)}{g(\vec x, y)} \, \sigma \, 0 \right],
\] 
and thus the simplifications from \cite{Uncuetal2023a} are faithful to the domains of the rational functions.  In other words, we maintain the validity of (\ref{UDE-a}) and (\ref{UDE-b}).

We did not run into the same problems as Section \ref{subsec:CBProblem} because we considered only existential quantification.  

\subsection{Universal quantification and denominators}

Now consider a rational function constraint under universal quantification: 
\[
\forall y \left[ \frac{f(\vec x,y)}{g(\vec x, y)} \, \sigma \, 0 \right].
\]
As outlined in Section \ref{subsec:restricted_domains} this should actually be read as 
\[
\forall y \in\{z : g(\vec x,z)\not= 0)\} \left[ \frac{f(\vec x,y)}{g(\vec x, y)} \, \sigma \, 0 \right].
\] 
Then its logical equivalence with an unconditional quantification must use implication to avoid making a statement when the denominator vanishes: 
\begin{align*}
&\forall y \left[ g(\vec x,y)\not= 0 \Rightarrow \frac{f(\vec x,y)}{g(\vec x, y)} \, \sigma \, 0\right],  \\
\Longleftrightarrow \, &\forall y \left[ \lnot \left( g(\vec x,y) \neq 0 \right) \lor \frac{f(\vec x,y)}{g(\vec x, y)} \, \sigma \, 0 \right], \\
\Longleftrightarrow \, &\forall y \left[ g(\vec x,y)= 0 \lor \frac{f(\vec x,y)}{g(\vec x, y)} \, \sigma \, 0 \right].
\end{align*} 
This then can be rewritten as
\begin{align}
\forall y \left[ g(\vec x,y)= 0 \lor \frac{f(\vec x,y)}{g(\vec x, y)} \, \sigma \, 0 \right]&\Longleftrightarrow\forall y \left[ g(\vec x,y)= 0 \lor {f(\vec x,y)}{g(\vec x, y)} \, \sigma \, 0 \right]
\end{align}
and
\begin{align}
\forall y \left[ g(\vec x,y)= 0 \lor \frac{f(\vec x,y)}{g(\vec x, y)} \, \sigma \, 0 \right]&\Longleftrightarrow\forall y \left[ g(\vec x,y)= 0\ \lor \right. \\\nonumber &\quad\quad \left.(g(\vec x, y)>0 \land f(\vec x,y) \, \sigma \, 0) \label{UDE-b} 
 \lor (g(\vec x, y)<0 \land 0 \, \sigma \, f(\vec x,y) )\right]
\end{align}
analogous to \eqref{UDE-a} and \eqref{UDE-b}, respectively.

\subsection{Solving the problem in Section \ref{subsec:CBProblem}}

To finish this section let us readdress the example introduced in Section \ref{subsec:CBProblem}.  In the first step below we make explicit the restricted domain and in the second we apply the universal quantification equivalent of (\ref{UDE-a}), the first of the two approaches to clearing the denominator from \cite{Uncuetal2023a}.
\begin{align*}
\forall x \left[\frac{1}{x^2} \geq 0 \right]
&\Longleftrightarrow 
\forall x \left[ {x^2}\not=0 \Rightarrow \frac{1}{x^2} \geq 0 \right] \\ 
&\Longleftrightarrow \forall x \left[ \lnot({x^2}\not=0) \lor x^2 \geq 0 \right]\\
&\Longleftrightarrow \forall x \left[ {x^2}=0 \lor x^2 \geq 0 \right].
\end{align*}
The final line can be clearly seen as true for every $x\in\mathbb{R}$.  If we consider the negation of the original example then we analyse this as in Section \ref{subsec:CBProblem} and find it false.  This time the problem and its negation are not the same.  

If we had followed instead the universal equivalent of (\ref{UDE-b}), we would get
\begin{align*}
\forall x \left[\frac{1}{x^2} \geq 0 \right]
&\Longleftrightarrow \forall x \left[ {x^2}\not=0 \Rightarrow \frac{1}{x^2} \geq 0\right] \\ 
&\Longleftrightarrow \forall x \left[ \lnot({x^2}\not=0) \lor \left( (x^2>0 \land 1\geq0)\lor(x^2<0 \land 1\leq0)\right) \right]\\
&\Longleftrightarrow \forall x \left[ {x^2}=0 \lor (x^2 > 0 \land 1\geq 0) \right],
\end{align*}
which is also true for all $x \in \mathbb{R}$.

\section{Future Work}

\subsection{Iterated resultants for CAD}

The work in Section \ref{sec:discard} looked only at the resultants, not the discriminants, and indeed only at resultants of resultants. Undoubtedly something similar could be said about, for example,
\begin{equation}\label{eq:disc}
\res_y(\res_z(f,g),\disc_z(f)).
\end{equation} 
We may observe that, in the case of the polynomials from Section \ref{subsec:detecting}, the evaluation of (\ref{eq:disc}) has a non-trivial square-free decomposition: in this case a polynomial of degree 58 times the square of another polynomial of degree 58.    
In fact, there is a known generic
factorization theorem pertaining to (\ref{eq:disc}):
generically, the square of $\res_{y,z}(f, g, \partial f/\partial z)$
is a factor of (\ref{eq:disc}) \cite{McCallum1999b, BuseMourrain2009}.
We would need a complete solution, applicable in the generic and non-generic cases, for the construction of (\ref{eq:disc}), and for resultants of discriminants, discriminants of resultants, discriminants of discriminants, etc., in order to firmly conclude a reduction in the worst-case degree growth.


Further, if we look again at (\ref{eq:4}) from the example in Section \ref{sec:Factorisation}, we see that this polynomial, which is the ``genuine'' part, factors further, and one factor has no real roots. Hence this factor can be discarded, though there is not much benefit, since we are at the univariate phase. Nevertheless, this shows that even the ``genuine'' part may still be overkill for \emph{real} geometry. Can we
\begin{enumerate}
\item[a)] detect that a factor of a resultant etc. has no real components; and
\item[b)]use this to further reduce the polynomials? Furthermore,
\item[c)]can we make any meaningful statement about the complexity implications of this?
\end{enumerate}

Another avenue for future work is to generalise the ideas of Sections \ref{sec:discard} and \ref{sec:MB} beyond the case of two or three equational constraints, moving toward a $k$-step projection in some sense where $k$ is informed by the number of equational constraints present.
Once that is done, theoretical and empirical comparison of the approaches of Sections \ref{sec:discard} and \ref{sec:MB} would be most interesting to carry out.

\subsection{Iterated resultants in CAD variants}
\label{subsec:CAC}

Over the past decade the theory of CAD has been used within a variety of other algorithms, e.g. NLSAT \cite{JovanovicdeMoura2012a}, NuCAD \cite{Brown2015a} and Cylindrical Algebraic Coverings \cite{Abrahametal2021a}, \cite{KremerNalbach2023a}.  These were developed for, or inspired by, the ideas in the satisfiability checking community.  It would be natural to ask whether the optimisations described here for CAD would also apply for these other algorithms.

With Cylindrical Algebraic Coverings \cite{Abrahametal2021a}, each polynomial has (at least one) explicit reason for being where it is in the computation. For example, $\res_{x_n}(f_1,f_2)$ might be in the computation because of a specific root $\alpha$, where it is the case for $x_{n-1}>\alpha$ (until the next point) the regions ruled out by $f_1$ and $f_2$ overlap, whereas for $x_{n-1}<\alpha$ we need a further reason to rule out regions. The same might be true of $\res_{x_n}(f_1,f_3)$, needed because of a specific root $\beta$. Then (\ref{eq:resres}) tracks where $\alpha$ and $\beta$ meet. Hence in this context we are interested only in genuine roots, and again we can replace (\ref{eq:resres}) by $\res(f_1,f_2,f_i)$.  Such ideas would need to be worked through precisely with an implementation of the coverings algorithm.


\subsection*{Acknowledgments}

JHD, ME and AKU are supported by the UK's EPSRC, via the DEWCAD Project,  \emph{Pushing Back the Doubly-Exponential Wall of Cylindrical Algebraic Decomposition} (grant numbers EP/T015713/1 and  EP/T015748/1), as was SMcC's visit to the UK to work with JHD and ME. AKU acknowledges the support of Austrian Science Fund (FWF) project P3401-N.

The authors are grateful to Jasper Nalbach whose questions about \cite{DavenportEngland2023a} led us to provide the clarification in Section \ref{subsec:simpleMres} and the new Section \ref{subsec:Issue}; and to Chris Brown and Zoltan Kov\'{a}cs whose conversation prompted Section \ref{sec:denominators}. We are also grateful to Gregory Sankaran, Tereso del~R\'{i}o and Amirhosein Sadeghi Manesh for useful conversations on equational constraints.

Finally, we express our gratitude to the anonymous referees whose comments greatly improved the final version of this paper.

\end{document}